\documentclass[twocolumn,showpacs,preprintnumbers,amsmath,amssymb]{revtex4}
\usepackage{graphicx}% Include figure files
\usepackage{dcolumn}% Align table columns on decimal point
\usepackage{bm}% bold math

\usepackage{color}
\usepackage[overlay,absolute]{textpos}
\definecolor{mygray}{gray}{0.4}
\definecolor{mylink}{rgb}{0.2, 0.2, 0.5}
\usepackage[colorlinks=true,urlcolor=mylink,dvipdfm,bookmarksnumbered=true,%
pdfborder={0 0 0},pdfpagemode=UseNone,pdfstartview=FitH]{hyperref}

\begin{document}

\begin{textblock*}{\textwidth}[0,0](19mm,11.5mm)
\footnotesize\noindent
\begin{minipage}{\textwidth}
\center
\textcolor{mygray}{Journal link:}
\href{http://dx.doi.org/10.1103/PhysRevA.76.043840}{http://dx.doi.org/10.1103/PhysRevA.76.043840}
\end{minipage}
\end{textblock*}

\begin{textblock*}{0.6\textwidth}[0,0](19mm,261mm)
\footnotesize\noindent
\begin{minipage}{\textwidth}
\textcolor{mygray}{Journal ref:} \href{http://dx.doi.org/10.1103/PhysRevA.76.043840}{A. E. B. Nielsen and K. M{\o}lmer, Phys.\ Rev.\ A \textbf{76}, 043840 (2007)}.\\
\textcolor{mygray}{Copyright (2007) by the American Physical Society.}
\end{minipage}
\end{textblock*}

\title{Transforming squeezed light into a
large amplitude coherent state superposition}

\author{Anne E. B. Nielsen and Klaus M{\o}lmer}
\affiliation{Lundbeck Foundation Theoretical Center for Quantum
System Research, Department of Physics and Astronomy, University of
Aarhus, DK-8000 \AA rhus C, Denmark}

\begin{abstract}
A quantum superposition of two coherent states of light with small
amplitude can be obtained by subtracting a photon from a squeezed
vacuum state. In experiments this preparation can be made
conditioned on the detection of a photon in the field from a
squeezed light source. We propose and analyze an extended
measurement strategy which allows generation of high fidelity
coherent state superpositions with larger amplitude.
\end{abstract}


\maketitle

\section{Introduction}

A Schr\"odinger cat state is a quantum superposition of
macroscopically distinguishable states, and in quantum optics the
term is traditionally used for superpositions of different
\emph{coherent states} such as the \emph{even} $(+)$ and \emph{odd}
$(-)$ cat states
\begin{equation}\label{cat}
|\psi_\textrm{cat}^\pm\rangle=\frac{1}{\sqrt{2(1\pm
e^{-2|\alpha|^2})}}\left(|\alpha\rangle\pm|-\alpha\rangle\right).
\end{equation}
The odd cat state populates only odd photon number states, while the
even cat state populates only even photon number states. A squeezed
vacuum state, generated, e.g., in a degenerate optical parametric
oscillator (OPO), is a superposition of even photon number states,
and it has been suggested \cite{dakna} to generate approximate odd
cat states by subtracting a photon from such a state, i.e., by
applying the photon annihilation operator $\hat{a}$ to a squeezed
vacuum. As discussed in Refs.\ \cite{dakna,kim,sasaki,molmer} the
resulting states are close to odd cat states when the cat size
$|\alpha|$ is small, because both the odd cat and the photon
subtracted squeezed vacuum state approach a single-photon state in
the limit of very small $|\alpha|$ and very little squeezing, and
for small amplitudes the degree of squeezing can be adjusted such
that the ratio between the $n=1$ and $n=3$ number state components
of the photon subtracted field match the ratio of the odd cat state.
If the scheme is modified appropriately, it is also possible to
approximately generate arbitrary superpositions of $|\alpha\rangle$
and $|-\alpha\rangle$ for small $|\alpha|$ as shown theoretically in
Ref.\ \cite{takeoka}.

Schr\"odinger cat states are interesting probes for quantum
mechanical behavior at the mesoscopic and macroscopic level, and
they may be used to investigate the role of decoherence in large
systems. They have potential applications for high precision
probing, and in particular they are useful resource states in
optical quantum computing proposals which make use of linear optics
and photon counting \cite{ralphcatcomp}. The application in quantum
computing is particularly interesting, because the encoding of
qubits in coherent states (and the resulting need for their
superposition states, i.e., the cat states) only requires relatively
small coherent state amplitudes to be sufficiently distinguishable
by homodyne detection. For more detailed discussions of single- and
two-qubit measurements in the coherent state and in the odd-even cat
state basis, see \cite{ralphcatcomp}, where it also follows that
resource cat states with amplitudes of about
$|\alpha|=\sqrt{6}\approx2.45$ or larger are sufficient for quantum
computing. Small cats have been generated experimentally using the
method described above \cite{wenger,ourjoumtsevcat,neergaard,wakui},
but the theoretically obtainable fidelity drops below $90\%$, when
$|\alpha|$ exceeds $1.9$, and other means have to be applied to make
larger cats with high fidelity. It has thus been proposed to
generate cats of larger size by combining smaller cats on beam
splitters \cite{jeong} or by amplifying cat states in an optical
parametric amplifier \cite{filip}. With a two-photon number state as
input, a recent experimental scheme for cat state preparation
conditioned on the outcome of homodyne detection was demonstrated
with $|\alpha|=\sqrt{2.6}$ \cite{naturecat}.

We suggest here an approach, which heralds the production of larger
cat states by a number of photo detection events. Dakna {\it et
al}.\ \cite{dakna} considered the states conditioned on multiple
photo detection events, and theory \cite{nm2} and experiments
\cite{ourjoumtsev2} have demonstrated the production of two-photon
states in a signal beam conditioned on the detection of two idler
photons from a nondegenerate OPO. Here we shall combine the field
from the degenerate OPO with a coherent state field prior to
counting of the photon numbers, as this allows us to effectively
produce states which are mathematically equivalent to the result of
applying operators of the form
$\hat{O}_A=\prod_{i=1}^A(\hat{a}-\beta_i)$ to a squeezed vacuum
state, where $\beta_i$, $i=1,\ldots,A$, are adjustable complex
numbers. Restricting the values of $\beta_i$ so that one vanishes
and the others occur in pairs $\pm\beta$ (for A odd), we can rewrite
the operator $\hat{O}_A$ as $\hat{O}_{2k+1}=
\left(\prod_{i=1}^k\left(\hat{a}^2-\beta_i^2\right)\right)\hat{a}$,
and the resulting state is a superposition of odd photon number
states, but we now have $k$ free complex parameters in addition to
the squeezing parameter, which may be chosen to match more closely
the number state amplitudes to the ones of the odd cat state.
Similarly, an approximate even cat state may be produced by applying
the operator $\hat{O}_{2k}=
\prod_{i=1}^k\left(\hat{a}^2-\beta_i^2\right)$ to a squeezed vacuum
state. Application of $\hat{O}_{2k+1}$ ($\hat{O}_{2k}$) involves
annihilation of $2k+1$ ($2k$) photons and since the probability to
obtain each annihilation in a real experiment is small, we shall
mainly be concerned with the case $k=1$ below. In Sec.\ II we
compute the cat state fidelities obtained for $k=1$ and compare the
results with the fidelities obtained for $k=0$. In Sec.\ III we
discuss how the operator $\hat{O}_{2k+1}$ may be approximately
realized experimentally, and Sec. IV concludes the paper.

\section{Cat state fidelity}

\begin{figure}
\begin{center}
\includegraphics*[viewport=22 12 392 298,width=0.95\columnwidth]{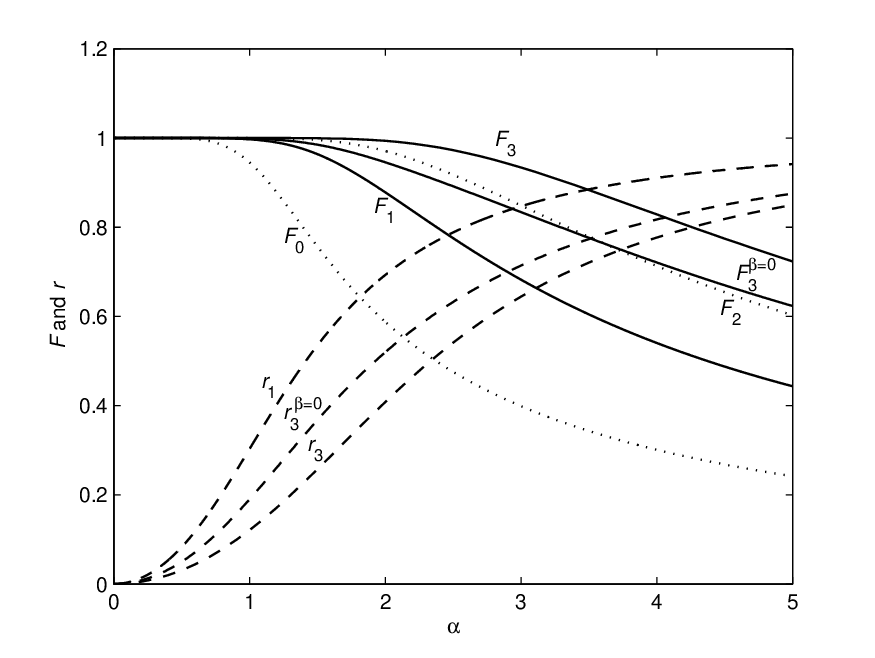}
\end{center}
\caption{The solid lines show the maximal odd cat state fidelity as
a function of $\alpha$ for annihilation of a single photon ($F_1$,
lower solid curve), annihilation of three photons ($F_3$, upper
solid curve), and annihilation of three photons when $\beta=0$ is
assumed ($F_3^{\beta=0}$, middle solid curve). The dashed lines are
the corresponding optimal values of $r$. The dotted lines give the
maximal even cat state fidelity for the squeezed vacuum state
($F_0$, lower dotted line) and for annihilation of two photons
($F_2$, upper dotted line)}\label{fid}
\end{figure}

We first determine the odd cat state fidelity for the case of a
single photo detection event ($k=0$). The initial single-mode
squeezed vacuum state may be expressed in the Fock state basis as
\cite{walls}
\begin{equation}\label{sq}
|\textrm{sq}\rangle=(1-r^2)^{1/4}
\sum_{n=0}^\infty\sqrt{\frac{(2n)!}{2^{2n}n!^2}}r^n|2n\rangle,
\end{equation}
where the squeezing parameter $r$, without loss of generality, is
assumed to be real and nonnegative. The variance of the squeezed
quadrature component in $|\textrm{sq}\rangle$ is reduced by the
factor $(1-r)/(1+r)$ (i.e., by
$-10\log_{10}\left((1-r)/(1+r)\right)\textrm{ dB}$) compared to the
vacuum value. The photon subtracted state is proportional to
$\hat{a}|\textrm{sq}\rangle$, and its overlap with an odd cat state
is
\begin{equation}
F_1=\frac{|\langle\psi_\textrm{cat}^-|\hat{a}|\textrm{sq}\rangle|^2}
{\langle\textrm{sq}|\hat{a}^\dag\hat{a}|\textrm{sq}\rangle}
=\frac{(1-r^2)^{3/2}|\alpha|^2e^{r\mathrm{Re}
(\alpha^2)}}{\sinh(|\alpha|^2)}.
\end{equation}
For a given desired cat size $|\alpha|$, the largest fidelity is
obtained for real $\alpha$ and
\begin{equation}
r=r_1\equiv\frac{\sqrt{9+4\alpha^4}-3}{2\alpha^2}.
\end{equation}
$r_1$ and the corresponding maximized fidelity are plotted as
functions of $\alpha$ in Fig.\ \ref{fid}. When the cat size
increases, the fidelity decreases and more squeezing is required.
For cat states with $|\alpha|>1.9$ the fidelity is below $0.9$, and
annihilation of a single photon from a squeezed vacuum state is thus
not a suitable method to generate large cats as stated in the
Introduction.

Turning now to the case of annihilation of three photons ($k=1$), we
find the odd cat state fidelity
\begin{eqnarray}\label{F3}
F_3&=&\frac{|\langle\psi_\textrm{cat}^-|
(\hat{a}^2-\beta^2)\hat{a}|\textrm{sq}\rangle|^2}{\langle
\textrm{sq}|\hat{a}^\dag((\hat{a}^\dag)^2-(\beta^*)^2)
(\hat{a}^2-\beta^2)\hat{a}|\textrm{sq}\rangle}\nonumber\\
&=&\frac{(3r+r^2(\alpha^*)^2-\beta^2)(3r+r^2\alpha^2-(\beta^*)^2)}
{|\beta|^4-3(\beta^2+(\beta^*)^2)\frac{r}{1-r^2}
+\frac{9r^2}{1-r^2}+\frac{15r^4}{(1-r^2)^2}}\nonumber\\
&&\frac{(1-r^2)^{3/2}|\alpha|^2e^{r\textrm{Re}(\alpha^2)}}
{\sinh(|\alpha|^2)}.
\end{eqnarray}
Since we have assumed that $r$ is real, it is optimal to choose
$\alpha$ real, and in this case the fidelity is maximized for
\begin{equation}\label{r3}
r=r_3\equiv\frac{\sqrt{\left(5+\sqrt{10}\right)^2+4\alpha^4}
-\left(5+\sqrt{10}\right)}{2\alpha^2}
\end{equation}
and
\begin{equation}\label{beta}
\beta^2=\beta_\textrm{opt}^2\equiv\frac{3}{7+2\sqrt{10}}\alpha^2.
\end{equation}
A second local maximum exists for $r$ and $\beta^2$ given by Eqs.\
\eqref{r3} and \eqref{beta} with $\sqrt{10}$ replaced by
$-\sqrt{10}$. The optimized fidelity and $r_3$ are plotted in Fig.\
\ref{fid}. The cat size, at which the fidelity drops below $0.90$,
is increased to $\alpha=3.3$, and the requested value of the degree
of squeezing is decreased significantly compared to the case of a
single count event. For $\alpha=\sqrt{6}$ we obtain the fidelity
$0.976$ for the optimal squeezing $r_3=0.53$ (i.e., 5.1 dB
squeezing, which is an experimentally realistic value \cite{lam}).

\begin{figure}
\begin{flushleft}
a)
\end{flushleft}
\begin{center}
\includegraphics*[viewport=17 8 392 298,width=0.95\columnwidth]{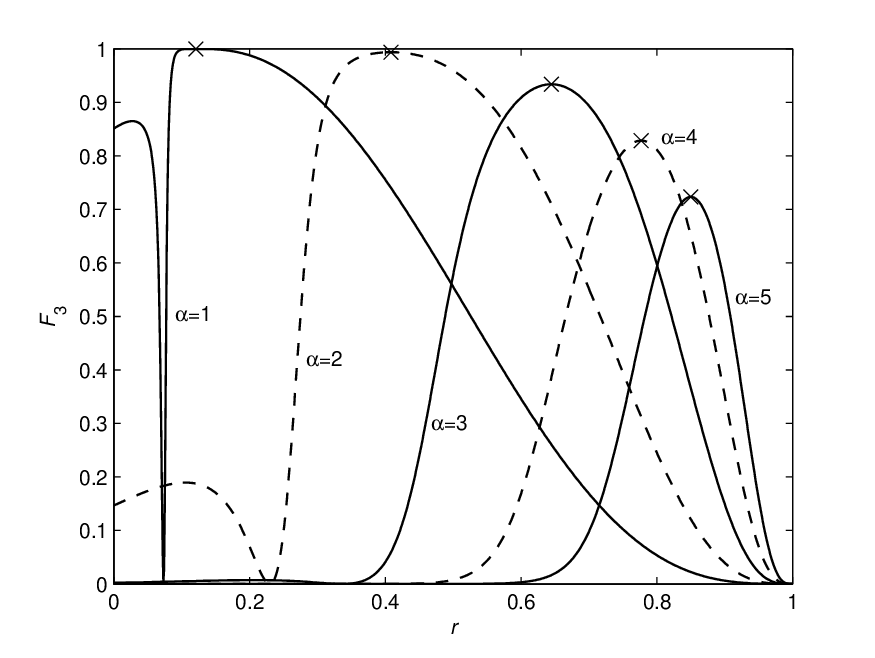}
\end{center}
\begin{flushleft}
b)
\end{flushleft}
\begin{center}
\includegraphics*[viewport=17 8 392 298,width=0.95\columnwidth]{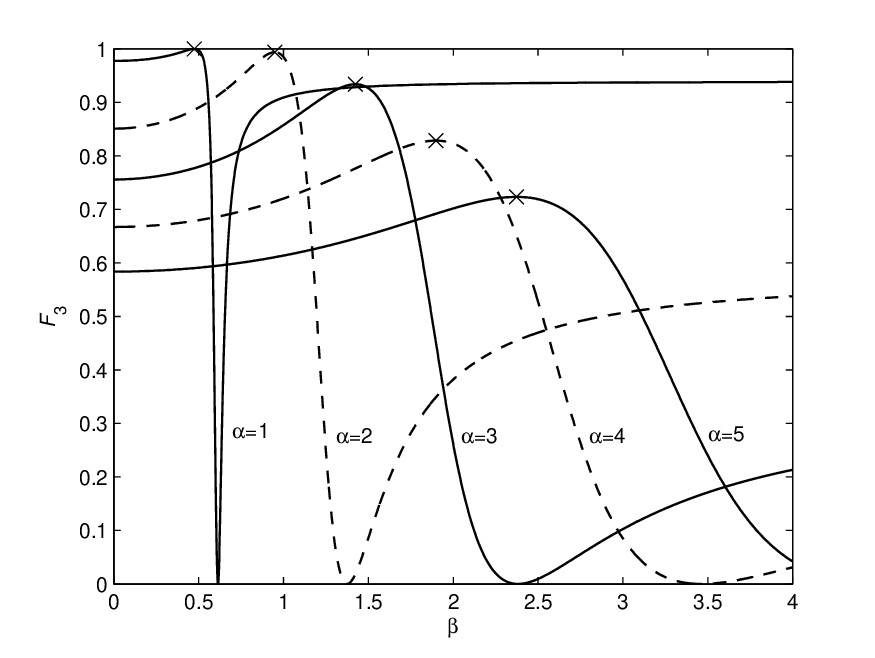}
\end{center}
\caption{Odd cat state fidelity for $2k+1=3$ (a) as a function of
$r$ for $\beta=\beta_{\textrm{opt}}$ and (b) as a function of
$\beta$ for $r=r_3$ for $\alpha=1,2,3,4,5$. The maximal fidelity is
marked with a cross on each line.} \label{tol}
\end{figure}

Figure \ref{tol} shows the fidelity as a function of $r$ for
$\beta=\beta_{\textrm{opt}}$ and as a function of $\beta$ for
$r=r_3$ for a few values of $\alpha$. This illustrates the
consequences of small deviations from the optimal values of the
parameters. It is, for instance, apparent that the parameters
leading to $3r+r^2\alpha^2-\beta^2=0$ and thus $F_3=0$ differ less
from the optimal parameters for small values of $\alpha$ than for
large values of $\alpha$, and the obtained fidelity may thus be more
sensitive to small deviations from the optimal parameters for small
values of $\alpha$, depending on the particular direction of the
change. It is, however, much simpler to generate these high
fidelity, small amplitude cats by subtracting only a single photon.

The experimental setup is less complicated if $\beta=0$, and Fig.\
\ref{fid} shows that also in this case the maximal fidelity is
increased compared to $k=0$. For our reference cat size,
$|\alpha|=\sqrt{6}$, the $\beta=0$ fidelity yields
$F_3^{\beta=0}=0.90$ for the squeezing parameter
$r_3^{\beta=0}=0.62$ (i.e., 6.3 dB squeezing).

Similar equations may be derived for even cat states, and the
resulting maximal fidelities are also plotted in Fig.\ \ref{fid}. It
is apparent that the maximal odd cat fidelity following application
of the operator $\hat{O}_3$, involving three annihilations, is
larger than the maximal even cat state fidelity following the
application of the operator $\hat{O}_2$, involving two
annihilations, and we thus focus on odd cat state generation in the
next section.

\section{Experimental implementation}

\begin{figure}
\begin{center}
\includegraphics*[viewport=53 106 345 266,width=0.85\columnwidth]{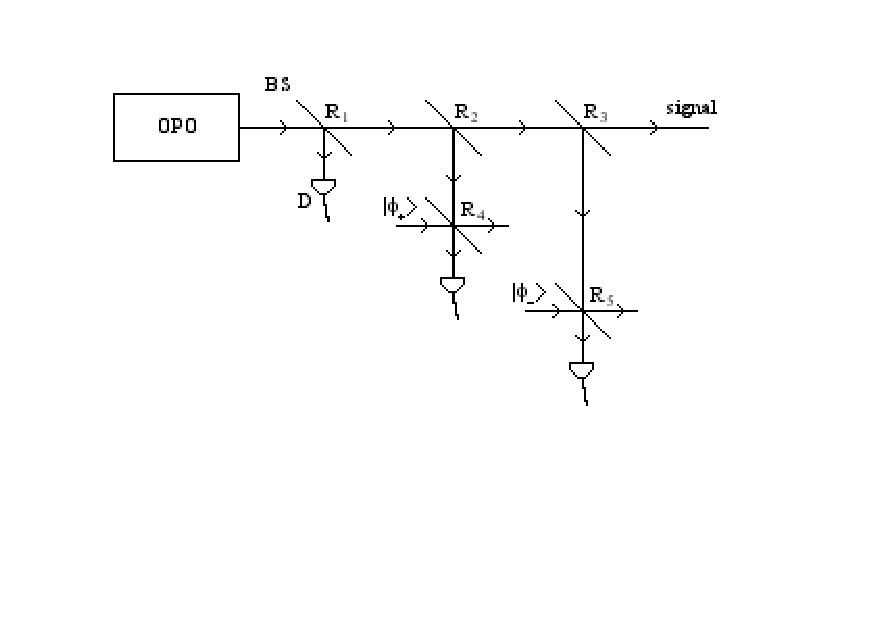}
\end{center}
\caption{Proposed experimental setup for generation of cat states
conditioned on three photo detection events. OPO, pulsed degenerate
optical parametric oscillator; BS, beam splitter; D, photo detector.
$R_i\equiv1-T_i<<1$ is the reflectivity of beam splitter $i$, and
$|\phi_\pm\rangle$ are input coherent states with amplitudes
$\phi_+=\sqrt{R_2/R_4}\beta$ and $\phi_-=-\sqrt{R_3/R_5}\beta$.}
\label{setup}
\end{figure}

The operator $\hat{O}_{2k+1}$ may ideally be implemented as shown
for $2k+1=3$ in Fig.\ \ref{setup}. The first beam splitter sends a
negligible small fraction of the initial squeezed vacuum state,
generated by the OPO, onto a photo detector, and application of the
first annihilation operator is obtained by conditioning on a photo
detection event. The experiments for $2k+1=1$ mentioned in the
Introduction also use this method. The operator
$\hat{a}^2-\beta_i^2$ can be rewritten as
$\hat{D}_{\hat{a}}^{-1}(-\beta_i)\hat{a}\hat{D}_{\hat{a}}(-\beta_i)
\hat{D}_{\hat{a}}^{-1}(\beta_i)\hat{a}\hat{D}_{\hat{a}}(\beta_i)$,
where
$\hat{D}_{\hat{a}}(\beta_i)=e^{\beta_i\hat{a}^\dag-\beta_i^*\hat{a}}$
is the field displacement operator \cite{walls}, which may be
implemented by mixing the state with a strong coherent state on a
beam splitter with a very small reflectivity \cite{paris}. To see
this, we imagine a beam splitter with transmission $\tau$ and feed
one input port with a coherent state $|\phi\rangle$ and the other
with an arbitrary input state with density operator
$\hat{\rho}_\textrm{in}$. Let $\hat{a}$ denote the annihilation
operator of the mode occupied by the input state and $\hat{b}$ the
annihilation operator of the mode occupied by the coherent state.
The action of the beam splitter is then represented by the unitary
operator
\begin{equation}
\hat{U}_{\tau}=\exp\left(i\tan^{-1}\left(\sqrt{\frac{1-\tau}{\tau}}\right)
\left(\hat{a}^\dag\hat{b}+\hat{a}\hat{b}^\dag\right)\right),
\end{equation}
and, after tracing out the $\hat{b}$ mode, we obtain the output
state
\begin{eqnarray}\label{rho}
\hat{\rho}_\textrm{out}&=&\frac{1}{\pi}\iint
\langle\gamma|\hat{U}_\tau|\phi\rangle\hat{\rho}_\textrm{in}
\langle\phi|\hat{U}_\tau^\dag|\gamma\rangle d\gamma_rd\gamma_i\nonumber\\
&=&\sqrt{\tau}^{\hat{a}^\dag\hat{a}}e^{-\zeta^*\phi\hat{a}^\dag+
\tau\zeta\phi^*\hat{a}}\sum_{n=0}^\infty\frac{1}{n!}(1-\tau)^n\hat{a}^n
\hat{\rho}_\textrm{in}(\hat{a}^\dag)^n\nonumber\\
&&\hspace{3cm}e^{-\zeta\phi^*\hat{a}+
\tau\zeta^*\phi\hat{a}^\dag}\sqrt{\tau}^{\hat{a}^\dag\hat{a}},
\end{eqnarray}
where $\gamma_r\equiv\textrm{Re}(\gamma)$,
$\gamma_i\equiv\textrm{Im}(\gamma)$, $|\gamma\rangle$ is a coherent
state, and $\zeta\equiv i\sqrt{(1-\tau)/\tau}$. In the limit
$\sqrt{1-\tau}\rightarrow0$, $\phi\rightarrow\infty$, and
$\sqrt{1-\tau}\phi\rightarrow\textrm{constant}$, \eqref{rho} reduces
to
\begin{equation}
\hat{\rho}_\textrm{out}=\hat{D}_{\hat{a}}(i\sqrt{1-\tau}\phi)\hat{\rho}_\textrm{in}
\hat{D}_{\hat{a}}^\dag(i\sqrt{1-\tau}\phi).
\end{equation}

This analysis suggests that we separate a small fraction of the beam
for the first single photon detection, we then displace the
remaining field by the appropriate amplitude $\beta$ and extract a
fraction of that beam for photo detection, we subsequently displace
the field towards the opposite amplitude $-\beta$ and extract again
a small fraction for detection, and finally we apply a displacement
to the remaining field. If all three photo detection events occur,
the resulting state is given by the the desired
$\hat{O}_{3}|\textrm{sq}\rangle$. It is, however, not very efficient
first to displace the state, then annihilate a photon, and finally
displace it back again to obtain
$\hat{D}_{\hat{a}}^{-1}(\beta_i)\hat{a}\hat{D}_{\hat{a}}(\beta_i)$,
and we suggest instead to displace only the small fractions of the
field which are subject to photo detection. The inverse
displacements are then not necessary. If we start with the state
$\hat{\rho}_\textrm{in}$, subtract a small fraction using a beam
splitter with reflectivity $R=1-T$, displace this fraction by the
amount $i\sqrt{R}\beta_i$ according to the above procedure,
annihilate a photon in the displaced mode $\hat{b}$, and trace out
the detected mode, we obtain the state
\begin{equation}\label{rhoout}
\hat{\rho}_\textrm{out}\propto\sum_{n=0}^\infty\langle
n|\hat{b}\hat{D}_{\hat{b}}(i\sqrt{R}\beta_i)\hat{U}_T|0\rangle\hat{\rho}_\textrm{in}
\langle0|\hat{U}_T^\dag\hat{D}_{\hat{b}}^\dag(i\sqrt{R}\beta_i)\hat{b}^\dag|n\rangle
\end{equation}
for the transmitted field. For $R\ll1$, we may expand \eqref{rhoout}
in orders of $R$, and to lowest order we find
\begin{equation}\label{rhoout2}
\hat{\rho}_{\textrm{out}}\propto
R(\hat{a}+\beta_i)\rho_{\textrm{in}}(\hat{a}^\dag+\beta_i^*),
\end{equation}
which is precisely the desired result.

A drawback of this experimental implementation is that, in the ideal
limit of zero beam splitter reflectivities, the success probability,
i.e., the probability to obtain the trigger detection events in a
given pulse of the setup, vanishes. It is thus necessary to allow
nonzero reflectivities, but this compromises the desired output
\eqref{rhoout2}, as it effectively induces a loss from the output
channel into the trigger channel. Large cat states are very
sensitive to losses, and, in general, it is necessary to keep losses
at an absolute minimum in order to obtain high fidelity, large
amplitude cat states. To illustrate this point, we imagine the
effect of sending a unit fidelity odd cat state of size $|\alpha|$
through a beam splitter with reflectivity $R=1-T$. After the beam
splitter the overlap with a cat of size $|\alpha|$ is decreased to
\begin{multline}
\sum_{m=0}^\infty\langle m|\langle\psi_\textrm{cat}^-|\hat{U}_T
|\psi_\textrm{cat}^-\rangle|0\rangle\langle0|\langle\psi_\textrm{cat}^-|
\hat{U}_T^\dag|\psi_\textrm{cat}^-\rangle|m\rangle=\\
\frac{\cosh(R|\alpha|^2)\sinh^2(\sqrt{T}|\alpha|^2)}{\sinh^2(|\alpha|^2)},
\end{multline}
and for $|\alpha|^2=6$ and $R=0.01$ this expression evaluates to
$0.94$. In the following we show that it is possible to obtain
fidelities that are essentially equal to those given in Fig.\
\ref{fid} and simultaneously obtain acceptable success probabilities
if the standard photo detectors are replaced by single photon number
resolving detectors. The qualitative explanation for this for
$\beta=0$ is that larger beam splitter reflectivities are allowable
if the detectors are able to weed out the instances where more than
a single photon are reflected at one of the beam splitters. We note
that, if losses are negligible, a single photon detector may be
built from a large number of unit efficiency photo detectors as
explained in \cite{rohde}.

\begin{figure}
\begin{center}
\includegraphics*[viewport=7 12 392 298,width=0.95\columnwidth]{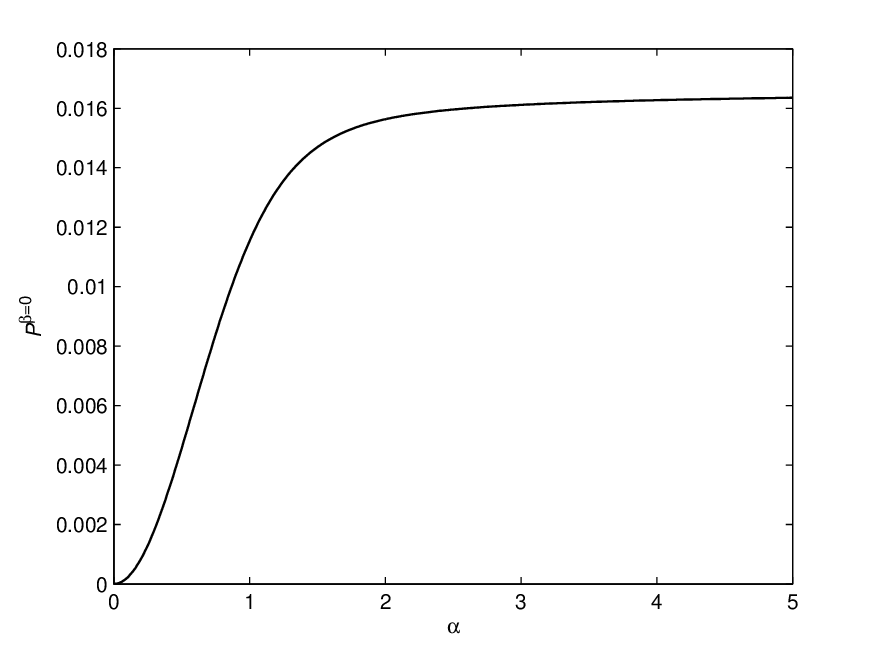}
\end{center}
\caption{Success probability as a function of $\alpha$ for the setup
with three single photon number resolving trigger detectors and
$\beta=0$. The success probability is maximized under the constraint
$x=r_3^{\beta=0}$ (see Fig.\ \ref{fid}).} \label{alfaP2}
\end{figure}

If the photo detectors in Fig.\ \ref{setup} are single photon number
resolving detectors, \eqref{rhoout} is replaced by
$|\psi_\textrm{out}\rangle\propto\langle1|\hat{D}_{\hat{b}}
(i\sqrt{R}\beta_i)\hat{U}_T|0\rangle|\psi_\textrm{in}\rangle$, and
the conditional, normalized output state following three single
photon detections is
\begin{multline}\label{psiout}
|\psi_\textrm{out}\rangle=\frac{1}{\sqrt{P}}
\langle1|\langle1|\langle1|
\hat{D}_{\hat{b}_3}(-ir_3\beta)\hat{U}_{T_3}\\
\hspace{3.5cm}\hat{D}_{\hat{b}_2}(ir_2\beta)
\hat{U}_{T_2}\hat{U}_{T_1}|0\rangle|0\rangle|0\rangle|\psi_\textrm{in}\rangle\\
=\frac{-i}{\sqrt{P}}\frac{r_1r_2r_3}{t_1t_2^2t_3^3}
e^{\frac{1}{2}(r_2^2+r_3^2)|\beta|^2}
\exp\left(\left(\frac{r_2^2}{t_2}-r_3^2\right)\frac{\beta^*}{t_3}\hat{a}\right)\\
(\hat{a}^2-t_2t_3^2\beta^2+(t_2-1)t_3\beta\hat{a})\hat{a}(t_3t_2t_1)
^{\hat{a}^\dag\hat{a}}|\psi_\textrm{in}\rangle,
\end{multline}
where $P$ is the success probability, $r_i=\sqrt{R_i}$,
$t_i=\sqrt{T_i}$, and $\hat{b}_1$, $\hat{b}_2$, and $\hat{b}_3$ are
annihilation operators of the three detected modes. Due to the
exponential factor in $\hat{a}$ and the term
$(t_2-1)t_3\beta\hat{a}$, the operator acting on the input state
$|\psi_\textrm{in}\rangle$ now also includes terms that annihilate
an even number of photons if $\beta\neq0$. For $\beta=0$ and
$|\psi_\textrm{in}\rangle=|\textrm{sq}\rangle$, on the other hand,
the fidelity is again given by Eq.\ \eqref{F3}, except that $r$ is
replaced by $x\equiv rT_1T_2T_3$. Choosing $x=r_3^{\beta=0}$ we thus
obtain the fidelities given by the middle solid curve in Fig.\
\ref{fid}, and the beam splitter reflectivities may be chosen in
order to maximize the success probability (see Eq.\ \eqref{P}
below). The results of such an optimization are shown in Fig.\
\ref{alfaP2}, and $P^{\beta=0}$ is seen to be approximately
$1.6\cdot10^{-2}$ for values of $\alpha$ around $\sqrt{6}$.

Despite the terms annihilating an even number of photons, the
fidelity may be increased by choosing an optimized nonzero value of
$\beta^2$. For $\beta\neq0$ we can get rid of the exponential factor
in \eqref{psiout}, if we choose $r_3^2=r_2^2/t_2$, and if $R_2$ is
not too large, the term $(t_2-1)t_3\beta\hat{a}$ has little effect.
For $r_3^2=r_2^2/t_2$ and
$|\psi_\textrm{in}\rangle=|\textrm{sq}\rangle$ the expression for
the odd cat state fidelity
\begin{multline}
F_3=\frac{(1-x^2)^{3/2}|\alpha|^2e^{x\textrm{Re}(\alpha^2)}}
{\sinh(|\alpha|^2)}(3x+x^2(\alpha^*)^2-t_2t_3^2\beta^2)\\
(3x+x^2\alpha^2-t_2t_3^2(\beta^*)^2)\Big/
\Big(t_2^2t_3^4|\beta|^4-t_2t_3^2(\beta^2+(\beta^*)^2)\frac{3x}{1-x^2}\\
+(t_2-1)^2t_3^2|\beta|^2\frac{1+2x^2}{1-x^2}
+\frac{9x^2}{1-x^2}+\frac{15x^4}{(1-x^2)^2}\Big).
\end{multline}
is almost identical to Eq.\ \eqref{F3}, but $r$ is replaced by $x$,
$\beta^2$ is replaced by $t_2t_3^2\beta^2$, and an extra term has
been added in the denominator. Since this term will not influence
the fidelity significantly for $t_2$ close to unity, we expect that
it is nearly optimal to choose $x$ and $t_2t_3^2\beta^2$ in
accordance with Eqs.\ \eqref{r3} and \eqref{beta}, respectively. If
we choose $T_2$ such that the extra term is a factor of $10^{-3}$
smaller than the sum of the rest of the terms in the denominator,
the fidelity is only decreased by approximately 0.1\% compared to
the fidelities represented by the upper solid curve in Fig.\
\ref{fid}. The last parameter $T_1$ may now be chosen in order to
maximize the success probability
\begin{multline}\label{P}
P=\langle\psi_\textrm{out}|\psi_\textrm{out}\rangle
=\frac{r_1^2r_2^2r_3^2}{t_1^2t_2^4t_3^6}e^{(r_2^2+r_3^2)|\beta|^2}
\sqrt{\frac{1-r^2}{1-x^2}}\\
\bigg(t_2^2t_3^4|\beta|^4\frac{x^2}{1-x^2}+(t_2-1)^2t_3^2|\beta|^2
\frac{x^2(1+2x^2)}{(1-x^2)^2}\\
-2t_2t_3^2\mathrm{Re}(\beta^2)
\frac{3x^3}{(1-x^2)^2}+\frac{3x^4(3+2x^2)}{(1-x^2)^3}\bigg).
\end{multline}
The optimal choice for fixed $x$, $T_2$, and $T_3$ is
\begin{equation}\label{T1}
T_1=\frac{\sqrt{x^4+8T_2^2T_3^2x^2}-x^2}{2T_2^2T_3^2}
\end{equation}
(provided \eqref{T1} leads to valid values of $T_1$ and $r$), and
for $\alpha=\sqrt{6}$ we find $P=6\cdot10^{-4}$. The price to pay
for the increase in fidelity is thus a more complicated setup and a
decrease in the success probability, but if the repetition rate of
the experiment is around $10^6\textrm{ s}^{-1}$ (see Ref.\
\cite{ourjoumtsev2}), it is still possible to produce of order
$10^3$ cat states per second.

The protocol suggested in Ref.\ \cite{jeong} combines two cat states
of the same size $|\alpha|$ on a beam splitter to obtain a
superposition of a cat state of size $\sqrt{2}|\alpha|$ at one of
the output ports and a vacuum state at the other or opposite. A
conditional measurement performed on one of the output beams
projects the other beam on the cat state. With a single
amplification, cats of size $|\alpha|=\sqrt{6}$ can thus be
generated from cats of size $|\alpha|=\sqrt{3}\approx1.73$. Figure
\ref{fid} reveals, however, that the maximal fidelity of the cat
states with $|\alpha|=\sqrt{3}$, generated by subtracting a single
photon from a squeezed vacuum state, is only $0.93$, and to achieve
a higher fidelity one may instead start from smaller cats with
larger fidelity and then amplify the states multiple times. With a
single photon number resolving detector, the success probability to
generate a cat state of size $|\alpha|=\sqrt{3/2}$ by detecting a
single photon is $P=0.13$, and for twofold amplification the
probability to generate the initial four cat states simultaneously
is thus $3\cdot10^{-4}$. Since the amplification protocol is itself
probabilistic, the total success probability is approximately one or
two orders of magnitude smaller (see \cite{jeong}).

\section{Conclusion}
The use of linear optics combined with measurements for quantum
computing was proposed in \cite{klm}, where single photon states had
to be provided as an online resource. As pointed out in
\cite{ralphcatcomp}, sufficiently strong coherent states have
several advantages over single photon states, but for qubit
implementation one must have a means to provide Schr\"odinger cat
states of a large enough amplitude to ensure that the two components
$|\alpha\rangle$ and $|-\alpha\rangle$ of the cat state wave
function are almost orthogonal.

In this paper we have suggested a protocol that is suitable to
generate such high fidelity cat states from the output of an OPO.
The states are heralded by three joint photo detection events. Using
normal APD photon counters, high fidelity states may only be
obtained with very small success probability, but with detectors
that can discriminate a single photon from higher photon numbers, we
predict quite acceptable production yields of high fidelity states.

\end{document}